# Transmutation of Nuclear Waste and the future MYRRHA Demonstrator

**Alex C. Mueller**

Institut National de Physique Nucléaire et de Physique des Particules, CNRS, Paris, France

E-mail: mueller@in2p3.fr

**Abstract.** While a considerable and world-wide growth of the nuclear share in the global energy mix is desirable for many reasons, there are also, in particular in the "old world" major objections. These are both concerns about safety, in particular in the wake of the Fukushima nuclear accident and concerns about the long-term burden that is constituted by the radiotoxic waste from the spent fuel. With regard to the second topic, the present contribution will outline the concept of Partitioning & Transmutation (P&T), as scientific and technological answer. Deployment of P&T may use dedicated "Transmuter" or "Burner" reactors, using a fast neutron spectrum. For the transmutation of waste with a large content (up to 50%) of (very long-lived) Minor Actinides, a sub-critical reactor, using an external neutron source is a most attractive solution. It is constituted by coupling a proton accelerator, a spallation target and a subcritical core. This promising new technology is named ADS, for accelerator-driven system. The present paper aims at a short introduction into the field that has been characterized by a high collaborative activity during the last decade in Europe, in order to focus, in its later part, on the MYRRHA project as the European ADS technology demonstrator.

## 1. Introductory remarks

Obviously, an important aspect of sustained growth of our planet is the long-term availability of energy resources and their environmental impact. It definitely constitutes a major ingredient in the struggle to guarantee peace and freedom for all of us.

In this context, current scenarii typically predict a doubling of the world's primary energy need for 2050. Yet, on the other hand, easy access to fossil energy will become more and more difficult together with the quest for an environmental policy, which would ensure a major reduction in emissions of green-house gases in order to combat their impact. However, in order to be based on scientific reasoning, a condition for real effectiveness, such a policy has to rely, for each energy-producing system, on a thorough scientific life-cycle analysis [1] of the total environmental impact.

Such studies (see, e.g. [2], [3]) conclude on the attractiveness of nuclear power with respect to its very small $CO_2$ (and other greenhouse gas) emissions. Reference [3], e.g., quotes a range between 10 – 130 $g/kWh_{elec}$ from various studies[1], substantially lower than that of any fossil-fuelled power technology for electricity production, and in favourable competition with other "renewable" technologies.

[1] Note that the work by Leeuwen and Smith [4], that gave the highest values, is generally criticized because of its underlying, debatable and sometime inconsistent, assumptions, e.g. by [5].

It is an established fact that the amount of $CO_2$ yearly avoided in the past by the use of nuclear power in Europe has is about 900 million tonnes (e.g. European Commission [6]), i.e. roughly equivalent to the present emissions of the whole transport sector. Not only may one express the fear that a pre-mature phase-out will of nuclear energy will strongly increase these emissions (because of strongly increased coal consumption as in Germany), but one also may advocate that it is rather unlikely to reach the 2020 EU $CO_2$ reduction target of 20% without a considerable increase of the nuclear share in the mix of the energy-generating systems. This for example supported by the fact that the present European ("EU27") $CO_2$ emission for 1 kWh of electrical energy is 356 $g/kWh_{el}$ whereas, e.g., this value is 447 $g/kWh_{el}$ for Germany and 89 $g/kWh_{el}$ for France, respectively [7].

Here is neither the place to develop and discuss such statements in more detail nor to contribute to the fervent debate that is presently taking place in the wake of the Fukushima nuclear accident.

Instead, the present paper is devoted to (only) address some issues on the long-term environmental burden of nuclear waste from the present-"generation-2" (and immediate-future "generation-3") reactors[2]. Indeed one counts 62 of such reactors under construction and 155 more projected. It is in this global context that the prospect of "Partitioning and Transmutation" that will be described in the next chapter could play an important role, noting in passing the fact that about 2500 tons of spent fuel are produced every year by the reactors of the European Union [8].

As final introductory remark I would like to point out the large international effort that is presently made for the development of "generation-4" nuclear reactors that aim to comply for the, say 2040, horizon in an ideal way to the criteria of sustainability[3], safety, reliability, and proliferation resistance.

## 2. The rationale for partitioning and transmutation

Partitioning and Transmutation of nuclear waste implements the principle of sustainable development in a rather general way: *separating out* of the spent fuel (partitioning) the radiotoxic components for *recycling* them (transmutation) in a way to minimize their toxicity and recover their contained energy in a useful way, in other words, minimizing potential health hazards while optimizing benefit for society.

Partitioning and Transmutation principally addresses the elements beyond Uranium that were generated (through a combination of successive neutron captures and radioactive decays, see fig.1 and its detailed caption) in a fission reactor, i.e. the actinides Neptunium, Plutonium, Americium and Curium[4]. Table 1 shows, for some of the relevant isotopes their half life and their accumulated amount for "typical spent fuel" from a 1000 $MW_{el}$ LWR, fuelled by Uraniumoxide, and discharging yearly 23 tons (of heavy elements). Recycling these actinides in a fast neutron spectrum will transmute them through fission in predominantly short-lived fission products; see for more details the subsequent section in this paper.

[2] These are mainly *light water reactors LWR,* which using (mostly) pressurized "natural" water for cooling and moderation.

[3] Sustainability means here in particular that the energy content of the (natural) Uranium and of the "breaded" heavier elements is fully exploited. This requires so-called "fast" (i.e. not moderated) *"generation-IV"* reactors, since the fission of these species is essentially induced by the fast part of the neutron spectrum. See section 3 for more and also some quantitative information.

[4] Note that it could also be of interest to minimize the amount of the (long-lived) fission products Iodine and Technetium.

Under the assumption of a reasonable separation efficiency (99.9% for U and Pu, 99% for the minor actinides), the amount of trans-uranium elements sent to a final disposal is reduced by more than 2 orders of magnitude by partitioning and transmutation. The benefit with regard to radio-toxicity has been highlighted e.g. in [8] comparing it to uranium mines as reference: whereas the waste from an open fuel cycle needs about $10^6$ y to reach the "natural" toxicity level, that value can be reached in a few hundred years by the fuel-cycle closure from partitioning and transmutation.

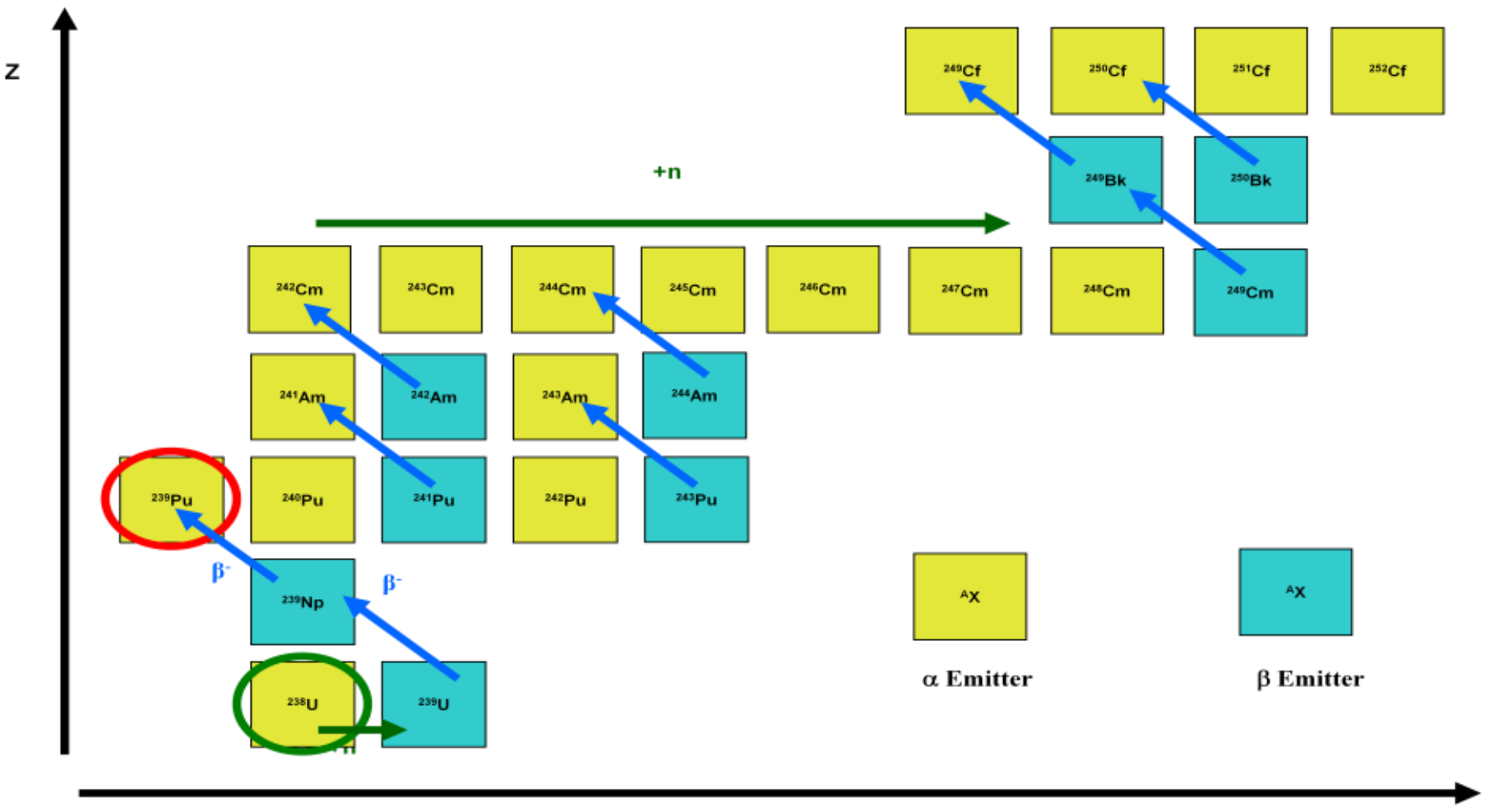

**Figure 1.** Neutrons from the fissioning isotope $^{235}$U of the fuel are captured, (e.g.), by the isotope $^{238}$U. The produced $^{239}$U will transform, trough two subsequent and rapid β-decays into $^{239}$Pu: $^{239}$U has half-life of 23 min and $^{239}$Np one of 2.4 days. Similarly, the isotopes of actinides with higher atomic number are synthesized. Note in passing that most countries will not consider the "breeded" $^{239}$Pu, fissionable in LWR (in contrast to most of other actinides) as waste, but as most valuable energy resource. Indeed, "partitioning" of $^{239}$Pu through the chemical reprocessing of the spent fuel, and adding it to fresh uranium fuel is the basis of the MOX (mixed-oxide) nuclear fuel-cycle depicted in figure 3b.

The substantial reduction of the actinides beyond Uranium strongly affects the *(long-term)* thermal behavior of the ultimate waste for a final disposal. The *initial* heat load is mainly due to the decay of the fission fragments Cesium and Strontium, which have rather short half lives, about 30 years. Thus an interim storage (or a separation of Cs and Sr) for hundred years, combined with partitioning and transmutation of the minor actinides would greatly enhance the capacity of a deep geological (i.e. granite or clay) repository, or, reasoning globally, considerably reduce the required number of sites for high-level waste by about an order of magnitude[5].

It has therefore been argued [10] that partitioning and transmutation is a necessary component for countries that will rely for the time to come on nuclear energy. Indeed, the availability of Uranium is stretched from the 100-200 year to several thousand year level, while simultaneously is reduced the amount of high-level radioactive waste per unit of energy generated. But further, partitioning and transmutation can also be highly useful for countries that envisage (an eventually gradual) phase-out of nuclear energy, since the burden of final storage is reduced both quantitatively as time-wise. From that, one can infer the potential of regional-level cooperation between countries with a different strategy with regard to nuclear power. This holds in particular true for an experimental scientific demonstration, as aimed by the MYRRHA project that I will address in a later chapter.

[5] Reference [9] quotes a factor of up to 50 for "optimal" conditions.

| **Isotope** | **Half live** | **Amount in kg/t** |
|---|---|---|
| $^{237}$Np | 2.1 My | 0.65 |
| $^{238}$Pu | 87.7 y | 0.23 |
| $^{239}$Pu | 24.1 Ky | 5.9 |
| $^{240}$Pu | 6.5 Ky | 2.6 |
| $^{241}$Pu | 14.35 y | 0.68 |
| $^{242}$Pu | 375 Ky | 0.6 |
| $^{241}$Am | 432 y | 0.77 |
| $^{243}$Am | 7.7 Ky | 0.14 |
| $^{244}$Cm | 18 y | 0.03 |

**Table 1.** Some typical long-lived heavy isotopes present in spent fuel (burn-up 40 GWd per ton of Uranium). The accumulated amount assumes an initial cooling-down time of 15 years, values from reference [8].

**3. Implementation of partitioning and transmutation**

Recently, within the EURATOM study PATEROS, several scenarii and different technical options for implementing partitioning and transmutation have been investigated to considerable detail [9]. Here we shall limit ourselves to some short recapitulation of the basic underlying physics and take the example of the, to our opinion particularly attractive, double-strata scenario with an accelerator-driven system ("ADS").

The transmutation physics is naturally driven by the underlying cross sections of the involved nuclear reactions; see as an example figure 2.

Derived from these, table 2 shows the so-called D-factor for the main nuclear species comparing moderated and fast neutron spectra. Its value describes the neutron consumption per fission. Correspondingly, a negative D-factor means that further neutrons are self-produced in excess, whereas a source of neutrons is required when $D \geq 0$ .

From the D-factor values in table 2 one immediately understands that transmutation requires the deployment of dedicated reactors with a fast neutron spectrum. In principle, both critical and sub-critical reactors are potential candidates as transmutation systems. *Fast critical reactors*, however, loaded with fuel containing *large amounts* of minor actinides (MA), namely Americium and Curium, pose problems caused by unfavourable reactivity coefficients and small delayed-neutron fractions. Hence it would be necessary to extend the fuel cycle for these minor actinides over a rather large ensemble of fast reactors rather than optimizing them for Plutonium burning and efficient electricity production.

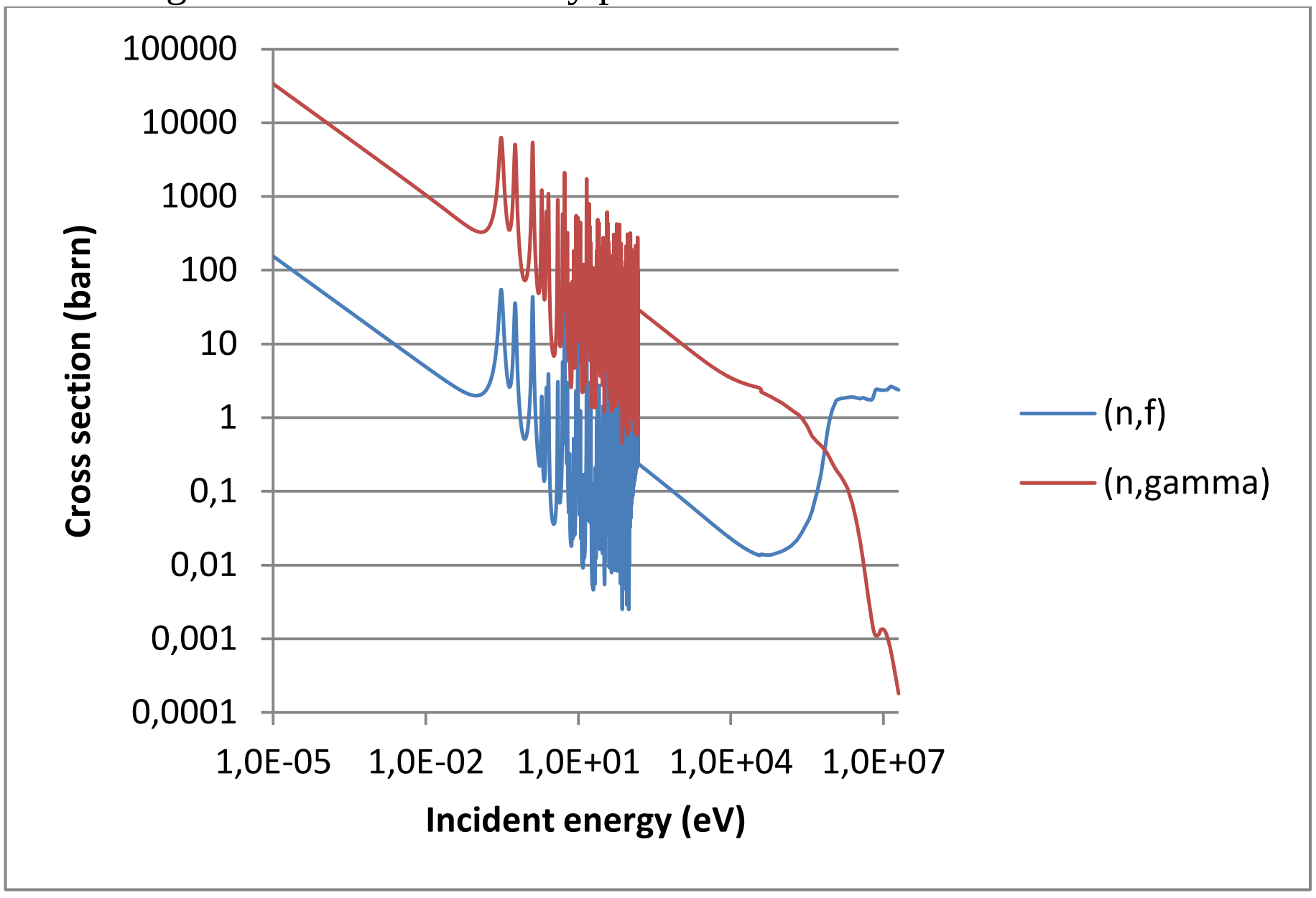

**Figure 2.** Neutron-capture and fission cross-sections as a function of the neutron energy for the example of the isotope $^{241}$Am. The neutron-capture is dominating until a neutron energy of about 1MeV by roughly 2 orders of magnitude. Thus heavier actinides (initially $^{242}$Am) are breeded, see also fig.1, by the neutrons from a moderated reactor. "Burning of the nuclear waste", that is transmutation into fission fragments, is only possible in a fast neutron reactor. Indeed, above 1 MeV, fission dominates with a constant cross section whereas capture eventually vanishes.

| Isotope | Fast Reactor | LWR |
|---|---|---|
| $^{238}$U | -0.62 | 0.07 |
| $^{238}$Pu | -1.36 | 0.17 |
| $^{239}$Pu | -1.46 | -0.67 |
| $^{240}$Pu | -0.96 | 0.44 |
| $^{241}$Pu | -1.24 | -0.56 |
| $^{242}$Pu | -0.44 | 1.76 |
| $^{237}$Np | -0.59 | 1.12 |
| $^{241}$Am | - 0.62 | 1.12 |
| $^{243}$Am | -0.60 | 0.82 |
| $^{244}$Cm | -1.39 | -0.15 |
| $^{245}$Cm | -2.15 | -1.48 |

**Table 2.** Comparison of the neutron consumption per fission for fast-spectrum and thermal (LWR) reactors from [8].

On the other hand, a *sub-critical system* using externally provided additional neutrons is here very attractive: it allows maximum (an order of magnitude higher) transmutation rates while operating in a safe manner. Coupling a proton accelerator, a spallation target and a sub-critical core, the name ADS, for Accelerator Driven System, is used for such a reactor. While technically feasible, it seems however less attractive (in particular also from an economic view) to deploy ADS technology all over the fuel cycle (i.e. to eventually replace *all* reactors by ADS).

Those arguments can justify a so-called *double-strata fuel cycle*. Electricity generation is performed in reactors with clean fresh fuel (only U and Pu), the present LWR being complemented (possibly phased out) by a gradual introduction of fast critical reactors. In the second stratum a (small) number of ADS is dedicated to the transmutation of minor actinides and any remaining Pu. Fig. 3 shows the different mentioned fuel-cycles in a pictorial way.

The development of ADS technology has been recently been addressed in various national and international research programs [see, e.g. 11, 12, 13]. The following section contains some selected information related to activities within EURATOM programmes and that focus today towards the timely construction of a European ADS-demonstrator.

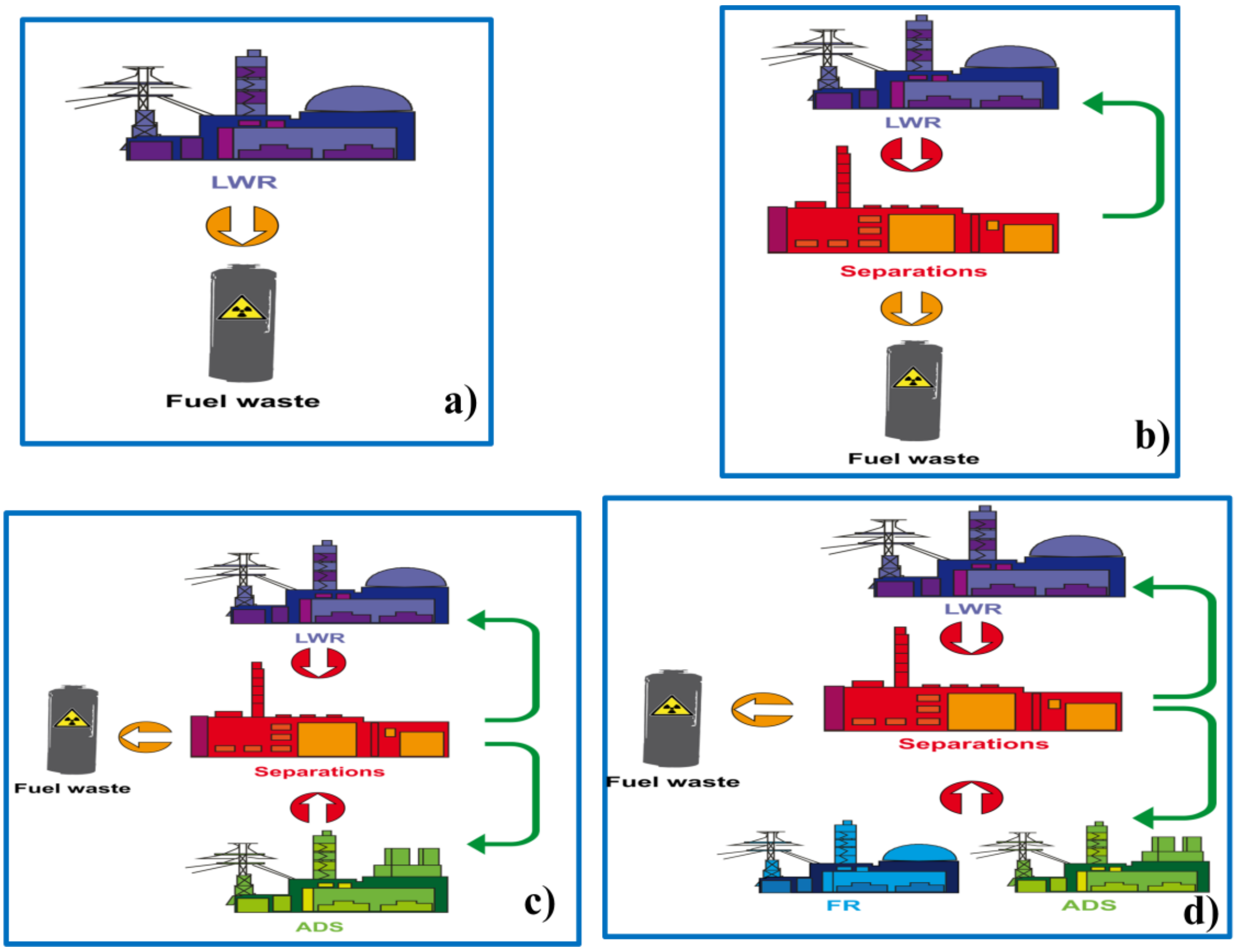

**Figure 3.** Different Fuel cycles as discussed in the text. a) corresponds to an open ("once-through") fuel cycle without out any reprocessing. b) recycling of the "partitioned" Plutonium

that is mixed with fresh Uranium fuel, such MOX-fuel is e.g. presently used in 16 reactors of the French Park. c) pictures an ADS deployment for transmuting the actinides left over in scenario b). d) is the "double-strata scenario" introducing fast critical ("generation4-") reactors that basically self-incinerate their own waste in scenario c).

## 4. Some selected information on European research programmes on ADS

The EUROpean research programme for the TRANSmutation of high-level nuclear waste in accelerator driven systems (EUROTRANS) was funded by the European Commission within the 6$^{th}$ EURATOM Framework Program (FP6). It involved 31 partners (research agencies and nuclear industries) with the contribution of 16 universities [11, 14, 15, 16]. EUROTRANS was a 5 year program (2005-2010) extending the previous FP5 "Preliminary Design Study for an eXperimental Accelerator Driven System" (PDS-XADS), described in reference [17]. One important aim of EUROTRANS was to pave the way towards the construction of an eXperimental facility demonstrating the technical feasibility of Transmutation in an Accelerator Driven System ("XT-ADS").

Within the EUROTRANS programme, the activities were split into five in technical areas (called Domains), respectively devoted to: the design of the ADS system and its sub-components, small-scale experiments on the coupling of an accelerator, a spallation target and a sub-critical core, studies on advanced fuels for transmuters, investigations on suited structural materials and heavy liquid metal technology, collection of nuclear data for transmutation. The main objective was to work towards a European Transmutation Demonstration in a step-wise manner, i.e. to provide (i) an advanced design of all the components of an XT-ADS system, (ii) a generic conceptual design of a modular European Facility for Industrial Transmutation (EFIT) for the long-term objective of the program.

The XT-ADS machine, which will essentially be loaded with conventional MOX fuel, is meant to be built and tested in the near future (next decade) so as to fulfil three main objectives. First, it should demonstrate the ADS concept (coupling of proton accelerator, spallation target and sub-critical assembly) at significantly high core power levels (50 to 100 $MW_{th}$). Secondly, it should validate the minor actinide (MA) transmutation by providing some dedicated positions in the core for special fuel assemblies. Finally, it should serve as a general multi-purpose neutron-irradiation facility. As such, it will also be an important asset for the qualification of different EFIT components.

| | **XT-ADS (ADS Prototype)** | **EFIT (Industrial Transmuter)** |
|---|---|---|
| **GOALS** | Demonstrate the concept<br>Demonstrate the transmutation<br>Provide an irradiation facility | Maximise the transmutation efficiency<br>Easiness of operation & maintenance<br>High level of availability |
| **MAIN FEATURES** | 50 – 100 $MW_{th}$ power | Several 100 $MW_{th}$ power |
| | Keff around 0.95 | Keff around 0.97 |
| | 600 MeV, 2.5 mA proton beam (back-up: 350 MeV, 5 mA) | 800 MeV, 20 mA proton beam |
| | Conventional MOX fuel | Minor Actinide fuel |
| | Lead-Bismuth Eutectic coolant & target | Lead coolant & target (back-up: gaz) |

**Table 3.** Baseline characteristics of the accelerator-driven systems XT-ADS and EFIT. $K_{eff}$ is the effective (global) neutron multiplication factor, also known as criticality factor which, for a subcritical system, naturally $K_{eff}$ is < 1.

The EFIT facility was defined as an industrial-scale transmutation demonstrator system, loaded with transmutation-dedicated fuel. Obviously, part of that demonstration is that it can be achieved in an economical way. Characteristics of EFIT are therefore an optimisation (i) for transmutation efficiency, (ii) for ease of operation and maintenance, and (iii) for a high level of availability. Despite these sometimes rather different objectives, an early outcome of the European studies [18, 19] has been that XT-ADS and EFIT require accelerators that share the same fundamental features. Both designs, see table 3, rely on a superconducting linear accelerator (linac) and liquid-metal technology for the spallation target and as core-coolant[6]. For the accelerator, the dominating reasons for the choice are the intrinsic potential for extreme reliability (for more information see subsequent sections) and for straight-forward up-grade to higher energy and beam power.

In the course of EUROTRANS, it has become obvious that the "generic XT-ADS" will see its practical realisation as the MYRRHA project at SCK-CEN in Mol, Belgium. More information on MYRRHA will be given in section 6.

After the success of EUROTRANS as one single integrated project delivering a global vision, the subsequent R&D program of the present FP7 is now targeting very specific urgent issues for building the XT-ADS. Among others[7] we underline here the Central Design Team contract, CDT, [20] that develops MYRRHA-FASTEF (see section 6), the contract for the Accelerator, MAX, [21] and FREYA [22], the one for the coupling experiments with GUINEVERE, see the final section.

## 5 Reference design for the accelerator, optimized for reliability

The European Transmutation Demonstration requires a high-power proton accelerator operating in CW mode, ranging from some MW (XT-ADS operation) up to several 10 MW (EFIT). The main beam specifications are shown in table 4. The extremely high reliability requirement (number of beam trips) can immediately be identified as the main technological challenge.

The reference design for the accelerator has been conceptually defined developed during the PDS-XADS program [18, 19]. It is based on the use of a superconducting linac, see figure 4.

| | **XT-ADS** | **EFIT** |
|---|---|---|
| **max. beam intensity** | 2.5 – 4 mA | 20 mA |
| **proton energy** | 600 MeV | 800 MeV |
| **beam entry** | Vertical from above | |
| **allowed beam trips (>1sec)** | < 5 per 3-month operation cycle | < 3 per year |
| **beam stability** | Energy: ±1%, Intensity: ±2%, Size: ±10% | |

[6] The XT-ADS relies on (eutectic) Lead-Bismuth because of the possibility of a lower working temperature, however, this solution is not deployable on a large scale for lack of material.

[7] Other relevant FP 7 contracts are, e.g., ANDES for nuclear data (with the corresponding Infrastructure contract ERINDA), ARCAS for ADS and fast-reactor comparison studies, and GETMAT for nuclear materials.

| | |
|---|---|
| **beam time structure** | CW, including zero-current periods (200μs), repeated at low rate |

**Table 4.** Main specifications for the proton beam. The listed requirements are for driving the technology-demonstrator XT-ADS compared to the industrial prototype EFIT.

This choice guarantees a very modular and upgradeable machine (same concept for prototype and industrial scale), an excellent potential for reliability, and a high RF-to-beam efficiency thanks to superconductivity (optimized operation cost). For the injector, an ECR source with a normal conducting RFQ is used up to, e.g. 3 MeV, followed by an energy booster section which uses normal conducting H-type DTL or/and superconducting CH-DTL structures up to a transition energy around 20 MeV. This low-energy part of the linac is duplicated in order to provide a "hot stand-by". From then on, a single, fully modular, superconducting linac, based on different RF structures (spoke, elliptical), accelerates the beam up to the final energy (350, 600, 1000 MeV…). Finally a doubly-achromatic beam line with a redundant beam scanning system transports the beam up to the spallation target and defines its fingerprint. This basic concept was further refined within the EUROTRANS, that also allowed to build and test first prototypical components. The present (accelerator dedicated) FP7 MAX [21] aims at a further level of demonstration allowing to go into the MYRRHA construction project.

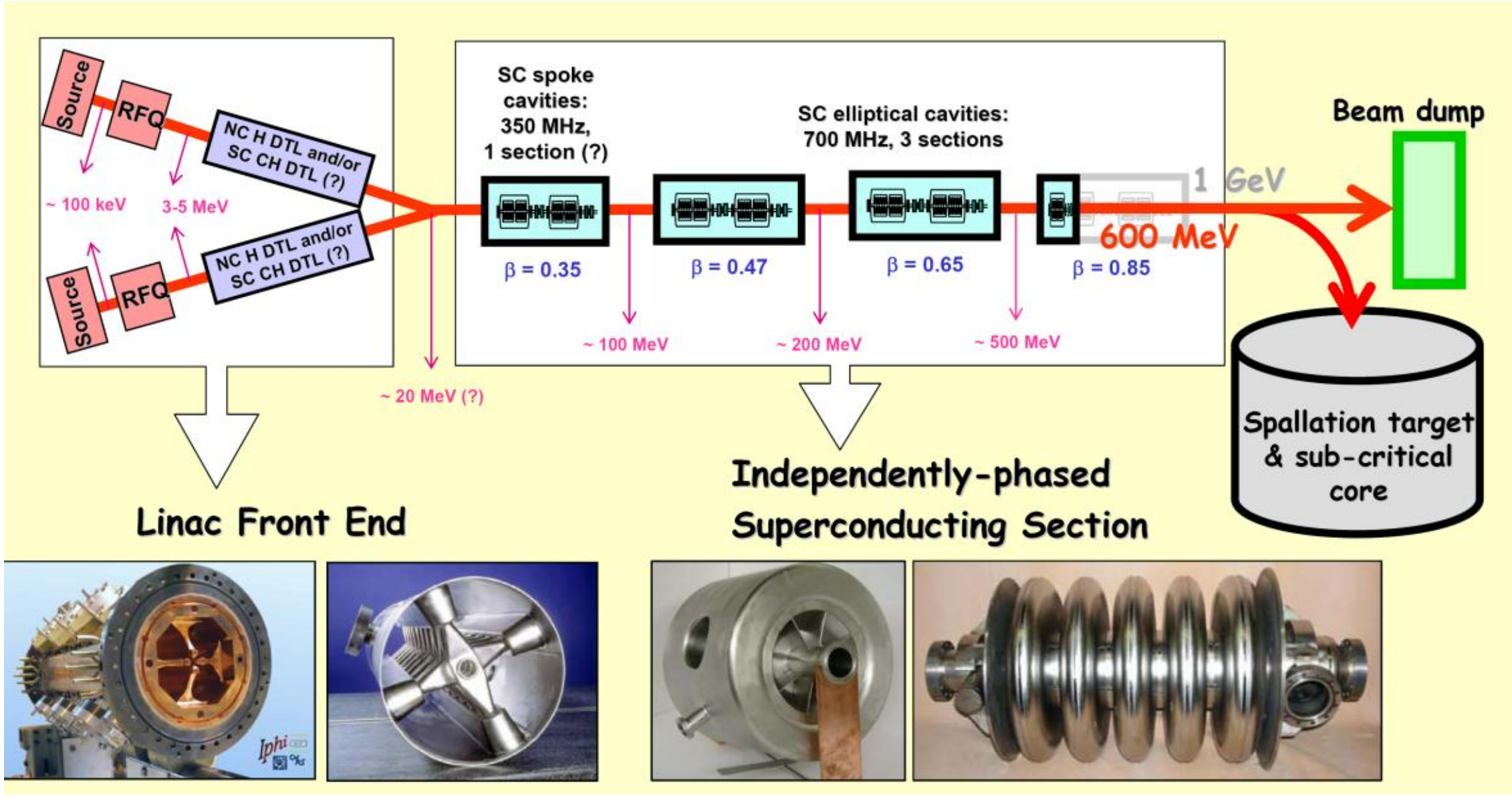

**Figure 4.** The reference accelerator scheme. The inserted photos show prototypes of the different RF structures that are deployed, from left to right: RFQ, CH-DTL, spoke- and elliptical cavity.

The ADS accelerator is expected – especially in running scenarii for the EFIT – to have a very limited number of unscheduled beam interruptions (per year!). Indeed, the consecutive interruption of the neutron production from the spallation target is of concern if it exceeds the second timescale.

This requirement is motivated by the fact that frequently repeated beam interruptions induce thermal stress and fatigue on the reactor structures, the target or the fuel elements, with possible significant damages, especially on the fuel claddings. Moreover, these beam interruptions decrease the plant availability, implying plant shut-downs in most of the cases. Therefore, it has been estimated that beam trips in excess of one second duration should not occur more frequently than five times per 3-month operation period for the XT-ADS, and three times per year for the EFIT.

To reach such an ambitious goal[8], which is lower than the reliability of the vast majority of accelerator based user facilities by 1-3 orders of magnitude[9], it is clear that reliability-oriented design practices were required since the conceptual design already, and they are also the main emphasize of the component design.

As a first and principle design rule, every linac component was conservatively de-rated with respect to its technological limitation *(over-design*). A high degree of *redundancy* has been planned, ab initio already, in critical areas. This is especially true for components where past engineering-experience has revealed limited reliability, e.g. linac injector and RF power systems[10]. A novel concept is the introduction of *fault-tolerance* wherever possible. These features can indeed be implemented to a very large extent in the highly modular superconducting RF linac above 20 MeV [24].

A preliminary bottom-up reliability analysis (Failure Mode and Effects Analysis, FMEA) has been performed in order to identify the critical areas in the design in terms of impact on the overall reliability [25]. It confirmed the choice of a second, redundant, 20 MeV proton injector (composed of the source, RFQ and low-energy booster), with fast switching capabilities.

After the injector stage, above 20 MeV, the superconducting linac forms an array of nearly identical “periods” in beam transport and acceleration. All components identically repeated, are modules that are operating well below any technological limitation. These two features allow the high degree of fault-tolerance for the accelerating cavities and the focusing magnets: neighbouring modules have enough "spare capacity" to assume temporarily the functions of a failing component. Necessarily, this approach implies a reliable and sophisticated machine control system. An important part of this is a digital RF control system for handling the RF set points in order to perform fast beam recovery in the case of cavity failures.

This basic concept was further refined within the EUROTRANS, that also allowed to build and test first prototypical components. The present (accelerator dedicated) FP7 MAX aims at a further level of demonstration allowing to go after its termination 2013 directly into the MYRRHA construction project. Summaries of all the recent developments for the accelerator have been given in [26, 27, 28, 29]

## 6. The MYRRHA project at SCK•CEN

The Belgian Nuclear Research Centre (SCK•CEN) in Mol has from the beginning been strongly involved in all the collaborative European projects in ADS technology. In fact, SCK•CEN was working since 1998 for its MYRRHA ADS-project, and was giving the MYRRHA technical design files as a starting point input for the EUROTRANS XT-ADS. MYRRHA, an acronym for Multipurpose hYbrid Research Reactor for High-tech

---

[8] Note that these values from EUROTRANS may be somewhat over-conservative as discussed presently within the CDT and MAX collaborations. Another re-assessment can be found in a recent study taking into account the operational experience from the PHOENIX fast reactor [23].

[9] Note that a typical state-of-the-art accelerator, namely the European Synchrotron Radiation Facility ESRF, very frequently runs one week without any interruption and even has recently reached the 4-week level according to private communications with Pascal Ellaume. Pascal, head of the accelerator division of ESRF tragically disappeared in a skiing accident in spring 2011, and the author wants to honor the memory of this outstanding accelerator physicist. Pascal was always very interested in news about the accelerator for MYRRHA and he was actually fully confident that the challenging reliability goals will be met.

[10] The increasing introduction of solid-state technology is of considerable interest in this context.

Applications; was planned as replacement for the MTR (material testing reactor) BR2 of the Mol research centre. Thus MYRRHA, while encompassing the role of XT-ADS is destined to allow an even wider nuclear technology R&D programme. A recent status report with a detailed description of the project can be found in [30, 31].

Indeed, MYRRHA is specified to have several irradiation stations, both in and around the reactor core. This guarantees the availability of a broad neutron spectrum, ranging from thermal energies to fast neutrons. Compared to classic MTRs, both the thermal and the fast neutron fluxes are very high, making it possible to simulate long-term exposure of materials in classical reactors. Moreover, the combination of high radiation damage (dpa) and the high production of H and He per unit of dpa close to the spallation target is particularly well suited for the study of materials for fusion applications. Thus the MYRRHA is devoted to the following tasks:

The first task consists in the demonstration of the complete ADS concept by coupling the three components (accelerator, spallation target and subcritical reactor) at a reasonably significant power level in order to allow operational feedback, scalable to an industrial demonstrator.

The second is the study of the efficient transmutation of high-level waste in dedicated places in the core.

The third is to be operated as a flexible irradiation facility. This will, i.a., allow a): fuel developments for innovative reactor systems; b) material developments for GEN IV systems and fusion reactors; c) Radioisotope production for medical and industrial applications; d) industrial irradiation applications such as Si-doping.

In order to fulfil this broad and ambitious program, and, moreover, to be intrinsically also a demonstrator of a Lead-cooled fast reactor, the EUROTRANS "MYRRHA XT-ADS" got a further refined design, "MYRRHA FASTEF" with the work accomplished within the FP7 contract CDT [20].

Thus one can consider that MYRRHA has now arrived in its Front-End Engineering Design Phase, ready for construction at the 2014 horizon, entering commissioning in 2020 and full-power operation by 2023, to name a few mile-stones on the MYRRHA road-map.

## 7. The coupling experiment GUINEVERE

A very significant step towards MYRRHA, is the GUINEVERE project [32]. GUINEVERE (**G**enerator of **U**ninterrupted **I**ntense **NE**utrons at the lead **VE**nus **RE**actor) was launched as of EUROTRANS, and now is continued within the FP7 project FREYA [22]. GUINEVERE is a zero power experiment for on-line reactivity monitoring and absolute reactivity measurements, both of which are major issues for ADS safety.

The VENUS reactor was used as zero-power thermal critical mock-up at SCK-CEN in Mol, Belgium until 2007, when it was modified changed into a lead fast reactor called VENUS-F[11] to be component of the GUINEVERE project. For this, VENUS-F is coupled to a neutron source driven by the GENEPI-3C deuteron accelerator, constructed by CNRS-IN2P3. This accelerator will not only be operated in pulsed mode, but also in continuous mode, the latter being the more representative of a powerful ADS. With that flexibility of the GENEPI accelerator VENUS-F will provide a unique facility in Europe where it will be possible to investigate both fast-critical and subcritical reactor. Figure 5 gives an overview of GUINEVERE.

---

[11]The fuel for the VENUS-F, provided by CEA, is metallic uranium (30% $^{235}U$ enrichment. See reference [31] for some details about the fuel assembly.

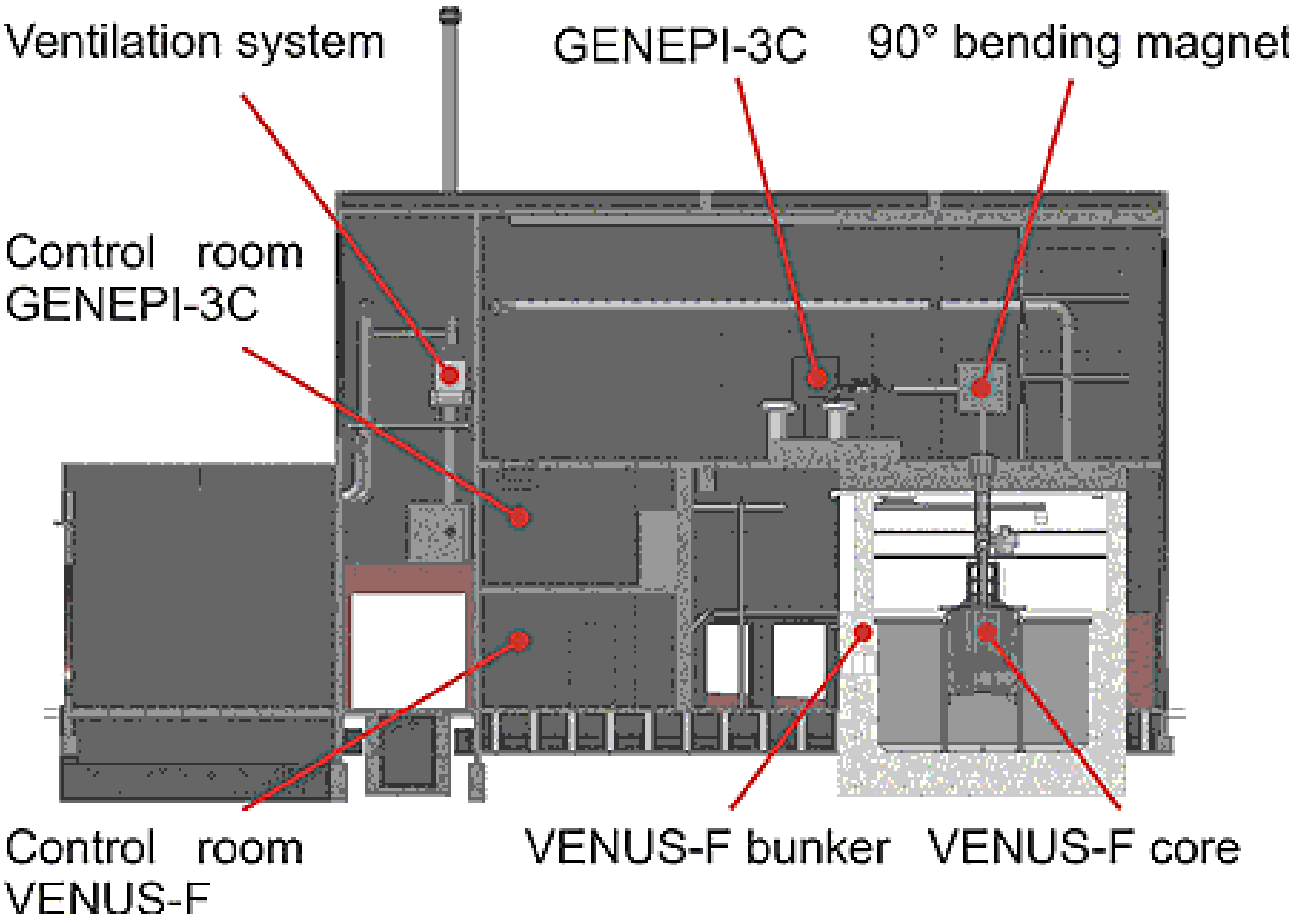

**Figure 5.** Side view of the VENUS facility modified to house the GUINEVERE experiments.

The installation is fully operational and the GUINEVERE coupling experiments are started. Thus one can be confident to rapidly obtain a substantial and consistent data base for ADS code validation that is important for licensing and operation of MYRRHA, e.g. like the relation between beam- and reactor-power as an on-line reactivity monitor. First such results have been published very recently [32].

## 8. Concluding remarks

The presented information (see in particular also the quoted references) intended to give a short introduction into the field of the closure of the nuclear fuel cycle, namely by Partitioning and Transmutation. The scientific and technical progress made in Europe within the different EURATOM projects allow to fix as next principal mile-stone the launch of the construction of the MYRRHA project in 2014 at its site in Mol, Belgium. Since MYRRHA is on the ESFRI-list of the future European Research Infrastructures and with the commitment of Belgium for an important part of the financing, one may hope that the formation of the international consortium that ensures the full funding of the project can accomplished in-line.

## 9. References

[1] Externalities and Energy Policy: *The Life Cycle Analysis Approach*, Proceedings of OECD-NEA Workshop, Paris, France, 15-16 November 2001, ISBN 92-64-18481-3
[2] A. Voss in [1], p. 163
[3] M. Bilek, M. Lenzen, C. Hardy and C. Dey, *Life-cycle energy balance and greenhouse gas emissions of nuclear energy in Australia*, Integrated Sustainability Analysis (ISA), University of Sydney (2006) http://www.isa.org.usyd.edu.au/publications/documents/ISA_Nuclear_Report.pdf
[4] J.W.S. van Leeuwen and P.B. Smith, *Nuclear Power – the Energy Balance* http://www.stormsmith.nl
[5] Commonwealth of Australia (2006), *Uranium Mining, Processing and Nuclear Energy* - ***Opportunities for Australia?*** **ISBN 0-9803115-0-0 978-0-9803115-0-1**

[6] Green paper by the European Commission COM (2006) 105: *A European Strategy for Sustainable, Competitive and Secure Energy* on Nuclear Energy {SEC(2006) 317}
[7] www.observatoire-electricite.fr
[8] C. Rubbia, H. A. Abderrahim, M. Björnberg, B. Carluec, G. Gherardi, E. Gonzalez Romero, W. Gudowski, G. Heusener, H. Leeb, W. von Lensa, G. Locatelli, J. Magill, J. M. Martínez-Val, S. Monti, A. C. Mueller, M. Napolitano, A. Pérez-Navarro, M.Salvatores, J. Carvalho Soares, J. B. Thomas, *The European Roadmap for Developing ADS for Nuclear Waste Incineration*, ISBN 88-8286-008-6, ENEA 2001
[9] E. Gonzales Romero, *P&T Rationale and added value for HLW management*, deliverable of the European Contract PATEROS, http://www.sckcen.be/pateros
[10] J.M. Martinez-Val, *European Roadmap for the deployment of P&T in a regional context , deliverable of the European Contract PATEROS*, http://www.sckcen.be/pateros
[11] J.U. Knebel, H.A. Abderrahim, L. Cinotti, F. Delage, C. Fazio, M. Giot, B. Giraud, E. Gonzalez, G. Granget, S. Monti, A.C. Mueller, *EUROTRANS: European Research Programme for the Transmutation of High Level Nuclear Waste in an Accelerator Driven System*, Ninth Nuclear Energy Information Exchange Meeting on Actinide and Fission Product Partitioning & Transmutation Nîmes, France, 25-29 September 2006, OECD-NEA 2007, ISBN 978-92-64-99030-2
[12] see, e.g., http://www.iaea.org/inisnkm/nkm/aws/fnss/ and links therein
[13] R. Sheffield, B.-C. Na, G. Lawrence, K. Pasamehmetoglu, P. Pierini, G. Olry, A. C. Mueller, P. Schuurmans, K. Van Tichelen, L. Cinotti, Accelerator and Spallation Target Technologies for ADS Applications, Nuclear Energy Agency, Nuclear Science Status Report (2005), ISBN 92-64-01056-4
[14] J.U. Knebel, H. A. Abderrahim, L. Cinotti, F. Delage, C. Fazio, M. Giot, B. Giraud, E. Gonzalez, G. Granget, S. Monti, Alex C. Mueller, *European Research Programme for the Transmutation of High Level Nuclear Waste in an Accelerator Driven System*, in Proceedings of 6th International Conference on EU Research and Training in Reactor Systems FISA-2006 ISBN 92-79-01214-2 European Communities 2006, and ftp://ftp.cordis.europa.eu/pub/fp6-euratom/docs/fisa2006_proceedings_eur-21231_en.pdf
[15] D. de Bruyn, H. A. Abderrahim, G. Rimpault, L. Mansani, M. Reale, A. C. Mueller, A. Guertin, J.L. Biarrotte, J. Wallenius, C. Angulo, A. Orden, A. Rolfe, D. Struwe, M. Schikorr, A. Woayne-Hune, C. Artioli, *Achievements and Lessons learnt within the domain1 "Design" of the integrated project EUROTRANS*, Proceedings 1st International Workshop on Technology and Components of Accelerator Driven Systems, TCADS, Karlsruhe, Germany, 15-17 March 2010
OECD Publishing DOI: 10.1787/9789264117297-EN
[16] EUROTRANS, Euratom FP 6 contract FI6W-CT-2005-516520, http://nuclear-server.ka.fzk.de/eurotrans
[17] B. Giraud, S. Ehster, G. Locatelli, L. Cinotti, L. Mansani, J. Pirson, X. Jardi, M.T. Dominguez, K. Peers, R. Sunderland, D. Coors, T. Abram, G. Granget, G. Rimpault, E. Gonzalez, A. C. Mueller, H. Klein, H. Wider, F. Bianchi, D. Struwe, P. Pierini, J. Cetnar, P. Coddington, A. Hogenbirk, B.R. Sehgal, H. Aït Abderahim, J. Martinez-Val, V. Moreau, P. Vaz, D. Vandeplassche, P. Phlippen, *Preliminary Design Study of an Experimental Accelerator-Driven System*, Euratom FP5 Contract FIKW CT-2001-00179 Final Report (2005), http://cordis.europa.eu/fp5-euratom/src/lib_finalreports
[18] A. C. Mueller, *The PDS-XADS Reference Accelerator and its Radioprotection Issues*, Radiation Protection Dosimetry **116**, 442 (2005) 442
[19] J.-L. Biarrotte, S. Bousson, T. Junquera A. C. Mueller, A. Olivier, *A reference accelerator scheme for ADS applications*, Nucl. Inst Methods **A562** , 565 (2006)
[20] CDT, Euratom FP7 contract FP7-232527
[21] MAX, Euratom FP7 contract FP7-269565
[22] FREYA, Euratom FP7 contract FP7-269665
[23] G. Rimpault, Ph. Dardé, F. Mellier, R. Dagan, M. Schikorr, A. Weisenburger, D. Maes, V. Sobolev, B. Arien, D. Lamberts, D. De Bruyn, A. C. Mueller, J.L. Biarrotte, *The issue of Beam Trips of the accelerator for efficient ADS operation*, submitted Nuclear Technology (2012)
[24] J-L. Biarrotte, M. Novati, P. Pierini, D. Uriot, *Beam dynamics studies for the fault tolerance assessment of the PDS-XADS linac design*, 4th OECD NEA International Workshop on Utilization and Reliability of HPPA, May 2004, Daejon, S. Korea, ISBN 92-64-01380-6
[25] P. Pierini, *ADS Reliability Activities in Europe*, in proceedings quoted in [23]
[26] see, e.g. 5th OECD NEA International Workshop on Utilization and Reliability of HPPA, May 2007, Mol, Belgium, ISBN 978-92-64-04478-4
[27] Jean-Luc Biarrotte and Alex C. Mueller, *Accelerator Reference Design For The European ADS Demonstrator*, Proceedings of the First International Workshop on Technology And Components of Accelerator-Driven Systems, TCADS, Karlsruhe, Germany 15-17 March 2010
OECD Publishing doi: 10.1787/9789264117297-en

[28] J.L. Biarrotte, A.C. Mueller, H. Klein, P. Pierini, D. Vandeplassche, *Accelerator reference design for the MYRRHA European ADS demonstrator*, Proceedings of LINAC 2010, September 2010, http://accelconf.web.cern.ch/accelconf/LINAC2010/html/author.htm, p.440

[29] J.L. Biarrotte, *High Power Hadron Accelerators: Applications In Support of Nuclear Energy*, Proccedings 15th International Conference on Emerging Nuclear Energy Systems, San Francisco, USA (2011) http://ipnweb.in2p3.fr/MAX/images/stories/downloads/icenes-2011_paper_biarrotte_final.pdf

[30] Hamid Aït Abderrahim, *MYRRHA an innovative and unique research facility*, 10th International Topical Meeting on Nuclear Applications of Accelerators, April, 3-7, 2011, Knoxville, Tennessee, USA, in press31.

[31] Hamid Aït Abderrahim, Peter Baeten, Didier De Bruyn, Rafael Fernandez, *MYRRHA – A multi- purpose fast spectrum research reactor*, Energy Conversion and Management, ISSN 0196-8904, http://www.sciencedirect.com/science/article/pii/S0196890412000982

[32] P. Baeten, H. A. Abderrahim, G. Bergmans, A. Kochetkov, W. Uyttenhove, D. Vandeplassche, F. Vermeersch, G. Vittiglio, G. Band, M. Baylac, A. Billebaud, D. Bondoux, J. Bouvier, S. Chabod, J.M. deConto, P. Dessagne, G. Gaudiot, J.M. Gautier, G. Heitz, M. Kerveno, B. Launé, F.R. Lecolley, J.L. Lecoué, N. Marie, Y. Merrer, A. Nuttin, D. Reynet, J.C. Steckmeyer, F. Mellier, *The GUINEVERE Project at the VENUS-F Facility*, Proceedings ENC 2010 European Nuclear Conference, 30 May - 2 June 2010, Barcelona, Spain http://www.euronuclear.org/events/enc/enc2010/index.htm

[32] H.E. Thyebault, P. Baeten, A. Billebaud, A. Chabod, A. Kochetkov, F.-R. Lecolley, J.-L. Lecouey, G. Lehaut, N. Marie, F. Mellier, W.Uyttenhove, G.Vittiglio, J. Wagemans, G. Ban, P. Dessagne, M. Kerveno, J.-C. Steckmeyer, *The GUINEVERE Experiment: first PNS measurements in a lead moderated sub-critical fast core*
International Congress on the Advances in Nuclear Power Plants (ICAPP'12), Chicago, USA (2012)
http://hal.in2p3.fr/in2p3-00722493